\documentclass[aps,prl,twocolumn,showpacs,superscriptaddress,groupedaddress]{
revtex4}
\usepackage{graphicx}  
\usepackage{dcolumn}   
\usepackage{bm}        
\usepackage{amssymb}   
\usepackage{comment}

\newcommand{\Btbw}      {\mbox{$\mathcal{B}(t\to Wb)$}}

\newcommand{\Gtbw}      {\mbox{$\Gamma(t\to Wb)$}}

\newcommand{\met}       {\mbox{\ensuremath{\slash\kern-.7emE_{T}}}}

\newcommand{\ttbar}     {\mbox{$t\bar{t}$}}

\newcommand{\ppbar}     {\mbox{$p\bar{p}$}}

\begin{document}

\hspace{5.2in} \mbox{FERMILAB-PUB-10/394-E}

\title{Determination of the width of the top quark}
\affiliation{Universidad de Buenos Aires, Buenos Aires, Argentina}
\affiliation{LAFEX, Centro Brasileiro de Pesquisas F{\'\i}sicas, Rio de Janeiro, Brazil}
\affiliation{Universidade do Estado do Rio de Janeiro, Rio de Janeiro, Brazil}
\affiliation{Universidade Federal do ABC, Santo Andr\'e, Brazil}
\affiliation{Instituto de F\'{\i}sica Te\'orica, Universidade Estadual Paulista, S\~ao Paulo, Brazil}
\affiliation{Simon Fraser University, Vancouver, British Columbia, and York University, Toronto, Ontario, Canada}
\affiliation{University of Science and Technology of China, Hefei, People's Republic of China}
\affiliation{Universidad de los Andes, Bogot\'{a}, Colombia}
\affiliation{Charles University, Faculty of Mathematics and Physics, Center for Particle Physics, Prague, Czech Republic}
\affiliation{Czech Technical University in Prague, Prague, Czech Republic}
\affiliation{Center for Particle Physics, Institute of Physics, Academy of Sciences of the Czech Republic, Prague, Czech Republic}
\affiliation{Universidad San Francisco de Quito, Quito, Ecuador}
\affiliation{LPC, Universit\'e Blaise Pascal, CNRS/IN2P3, Clermont, France}
\affiliation{LPSC, Universit\'e Joseph Fourier Grenoble 1, CNRS/IN2P3, Institut National Polytechnique de Grenoble, Grenoble, France}
\affiliation{CPPM, Aix-Marseille Universit\'e, CNRS/IN2P3, Marseille, France}
\affiliation{LAL, Universit\'e Paris-Sud, CNRS/IN2P3, Orsay, France}
\affiliation{LPNHE, Universit\'es Paris VI and VII, CNRS/IN2P3, Paris, France}
\affiliation{CEA, Irfu, SPP, Saclay, France}
\affiliation{IPHC, Universit\'e de Strasbourg, CNRS/IN2P3, Strasbourg, France}
\affiliation{IPNL, Universit\'e Lyon 1, CNRS/IN2P3, Villeurbanne, France and Universit\'e de Lyon, Lyon, France}
\affiliation{III. Physikalisches Institut A, RWTH Aachen University, Aachen, Germany}
\affiliation{Physikalisches Institut, Universit{\"a}t Freiburg, Freiburg, Germany}
\affiliation{II. Physikalisches Institut, Georg-August-Universit{\"a}t G\"ottingen, G\"ottingen, Germany}
\affiliation{Institut f{\"u}r Physik, Universit{\"a}t Mainz, Mainz, Germany}
\affiliation{Ludwig-Maximilians-Universit{\"a}t M{\"u}nchen, M{\"u}nchen, Germany}
\affiliation{Fachbereich Physik, Bergische  Universit{\"a}t Wuppertal, Wuppertal, Germany}
\affiliation{Panjab University, Chandigarh, India}
\affiliation{Delhi University, Delhi, India}
\affiliation{Tata Institute of Fundamental Research, Mumbai, India}
\affiliation{University College Dublin, Dublin, Ireland}
\affiliation{Korea Detector Laboratory, Korea University, Seoul, Korea}
\affiliation{CINVESTAV, Mexico City, Mexico}
\affiliation{FOM-Institute NIKHEF and University of Amsterdam/NIKHEF, Amsterdam, The Netherlands}
\affiliation{Radboud University Nijmegen/NIKHEF, Nijmegen, The Netherlands}
\affiliation{Joint Institute for Nuclear Research, Dubna, Russia}
\affiliation{Institute for Theoretical and Experimental Physics, Moscow, Russia}
\affiliation{Moscow State University, Moscow, Russia}
\affiliation{Institute for High Energy Physics, Protvino, Russia}
\affiliation{Petersburg Nuclear Physics Institute, St. Petersburg, Russia}
\affiliation{Stockholm University, Stockholm and Uppsala University, Uppsala, Sweden }
\affiliation{Lancaster University, Lancaster LA1 4YB, United Kingdom}
\affiliation{Imperial College London, London SW7 2AZ, United Kingdom}
\affiliation{The University of Manchester, Manchester M13 9PL, United Kingdom}
\affiliation{University of Arizona, Tucson, Arizona 85721, USA}
\affiliation{University of California Riverside, Riverside, California 92521, USA}
\affiliation{Florida State University, Tallahassee, Florida 32306, USA}
\affiliation{Fermi National Accelerator Laboratory, Batavia, Illinois 60510, USA}
\affiliation{University of Illinois at Chicago, Chicago, Illinois 60607, USA}
\affiliation{Northern Illinois University, DeKalb, Illinois 60115, USA}
\affiliation{Northwestern University, Evanston, Illinois 60208, USA}
\affiliation{Indiana University, Bloomington, Indiana 47405, USA}
\affiliation{Purdue University Calumet, Hammond, Indiana 46323, USA}
\affiliation{University of Notre Dame, Notre Dame, Indiana 46556, USA}
\affiliation{Iowa State University, Ames, Iowa 50011, USA}
\affiliation{University of Kansas, Lawrence, Kansas 66045, USA}
\affiliation{Kansas State University, Manhattan, Kansas 66506, USA}
\affiliation{Louisiana Tech University, Ruston, Louisiana 71272, USA}
\affiliation{University of Maryland, College Park, Maryland 20742, USA}
\affiliation{Boston University, Boston, Massachusetts 02215, USA}
\affiliation{Northeastern University, Boston, Massachusetts 02115, USA}
\affiliation{University of Michigan, Ann Arbor, Michigan 48109, USA}
\affiliation{Michigan State University, East Lansing, Michigan 48824, USA}
\affiliation{University of Mississippi, University, Mississippi 38677, USA}
\affiliation{University of Nebraska, Lincoln, Nebraska 68588, USA}
\affiliation{Rutgers University, Piscataway, New Jersey 08855, USA}
\affiliation{Princeton University, Princeton, New Jersey 08544, USA}
\affiliation{State University of New York, Buffalo, New York 14260, USA}
\affiliation{Columbia University, New York, New York 10027, USA}
\affiliation{University of Rochester, Rochester, New York 14627, USA}
\affiliation{State University of New York, Stony Brook, New York 11794, USA}
\affiliation{Brookhaven National Laboratory, Upton, New York 11973, USA}
\affiliation{Langston University, Langston, Oklahoma 73050, USA}
\affiliation{University of Oklahoma, Norman, Oklahoma 73019, USA}
\affiliation{Oklahoma State University, Stillwater, Oklahoma 74078, USA}
\affiliation{Brown University, Providence, Rhode Island 02912, USA}
\affiliation{University of Texas, Arlington, Texas 76019, USA}
\affiliation{Southern Methodist University, Dallas, Texas 75275, USA}
\affiliation{Rice University, Houston, Texas 77005, USA}
\affiliation{University of Virginia, Charlottesville, Virginia 22901, USA}
\affiliation{University of Washington, Seattle, Washington 98195, USA}
\author{V.M.~Abazov} \affiliation{Joint Institute for Nuclear Research, Dubna, Russia}
\author{B.~Abbott} \affiliation{University of Oklahoma, Norman, Oklahoma 73019, USA}
\author{M.~Abolins} \affiliation{Michigan State University, East Lansing, Michigan 48824, USA}
\author{B.S.~Acharya} \affiliation{Tata Institute of Fundamental Research, Mumbai, India}
\author{M.~Adams} \affiliation{University of Illinois at Chicago, Chicago, Illinois 60607, USA}
\author{T.~Adams} \affiliation{Florida State University, Tallahassee, Florida 32306, USA}
\author{G.D.~Alexeev} \affiliation{Joint Institute for Nuclear Research, Dubna, Russia}
\author{G.~Alkhazov} \affiliation{Petersburg Nuclear Physics Institute, St. Petersburg, Russia}
\author{A.~Alton$^{a}$} \affiliation{University of Michigan, Ann Arbor, Michigan 48109, USA}
\author{G.~Alverson} \affiliation{Northeastern University, Boston, Massachusetts 02115, USA}
% ...
%
%
\collaboration{The D0 Collaboration\footnote{with visitors from
$^{a}$Augustana College, Sioux Falls, SD, USA,
$^{b}$The University of Liverpool, Liverpool, UK,
$^{c}$SLAC, Menlo Park, CA, USA,
$^{d}$ICREA/IFAE, Barcelona, Spain,
$^{e}$Centro de Investigacion en Computacion - IPN, Mexico City, Mexico,
$^{f}$ECFM, Universidad Autonoma de Sinaloa, Culiac\'an, Mexico,
and 
$^{g}$Universit{\"a}t Bern, Bern, Switzerland.%
}} \noaffiliation
\vskip 0.25cm

\date{September 28, 2010}

\begin{abstract}

We extract the total width of the top quark, $\Gamma_t$, from the
partial decay width $\Gtbw$ measured using the $t$-channel cross section
for single top quark production and from the branching fraction $\Btbw$
measured in $\ttbar$ events using up to 2.3~fb$^{-1}$ of integrated
luminosity collected by the D0 Collaboration at the Tevatron \ppbar\
Collider. The result is $\Gamma_t = 1.99^{+0.69}_{-0.55}$~GeV, which
translates to a top-quark lifetime of $\tau_t =
(3.3^{+1.3}_{-0.9})\times10^{-25}$~s.
Assuming a high mass fourth generation $b'$ quark and unitarity of
the four-generation quark-mixing matrix, we set the first upper limit on
$|V_{tb'}| < 0.63$ at 95\% C.L. 

\end{abstract}

\maketitle

\newpage


The total width, or lifetime, of the top quark is a fundamental
property that has not been measured precisely so far.
The top quark, like other fermions in the Standard Model (SM), decays
through the electroweak interaction. But unlike $b$ and $c$
quarks, which form long-lived hadrons that can be observed through the
reconstruction of
displaced vertices in a tracking detector,
the top quark has an extremely short lifetime.

In the SM, the total decay width of the top quark, $\Gamma_t$, is
dominated by the partial decay width $\Gamma(t\to Wb)$ which,
at next-to-leading order (NLO) in Quantum Chromodynamics (QCD),
depends on the top quark mass $m_t$, the $W$ boson mass $M_W$, the $b$ quark
mass $m_b$, the Fermi coupling
constant $G_F$, the strong coupling constant $\alpha_s$ and the
strength of the left-handed $Wtb$ coupling,~$V_{tb}$.
Neglecting higher order electroweak corrections~\cite{Denner:1990ns} and terms of
order $m_b^2/m_t^2$, $\alpha_s^2$, and 
$(\alpha_s/\pi)M_W^2/m_t^2$, the 
partial width becomes~\cite{pdg_top_review}
\begin{eqnarray}
\label{eq:smwidth}
\nonumber\Gamma (t\to Wb) &=& \frac{G_Fm_t^3}{8\pi\sqrt{2}}\, |V_{tb}|^2 \
\!\!\left ( 1 - \frac{M_W^2}{m_t^2} \right )^2 \!\!\left (1 +
2\frac{M_W^2}{m_t^2} \right ) \times \\
&& \!\! \times  \left [ 1 - \frac{2\alpha_s}{3\pi}  
\!\!\left ( \frac{2\pi^2}{3} - \frac{5}{2} \right )
\right ] \ .
\end{eqnarray}
Setting 
$\alpha_s (M_Z)= 0.118$, $G_F = 1.16637\times 10^{-5}~{\rm GeV}^{-2}$, $M_W =
80.399$~GeV, $|V_{tb}|=1$, and $m_t=170$~GeV leads to ${\Gtbw_{\rm
SM}}=1.26$~GeV.
%
%
%
%
%
Equation~\ref{eq:smwidth} can be extended to include non-SM $Wtb$
couplings~\cite{Chen:2005vr}. 

The decay width of an unstable particle can be measured with
precision from its mass spectrum when the experimental resolution
is similar or smaller than the natural width of the particle. 
Because $\Gamma_t$ is far smaller than the experimental resolution,
the analysis of the invariant mass distribution yields only an upper
limit on $\Gamma_t$ that is limited by the 
uncertainty on the detector resolution. 
The first such direct upper bound of $\Gamma_t < 13.1$~GeV was set by CDF at 
95\% C.L.~\cite{Aaltonen:2008ir}. 

\begin{figure}[!h!tbp]
\includegraphics[width=2.0in]{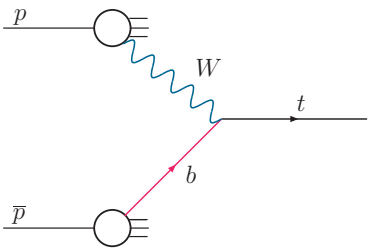}
\caption{Representative diagram for $t$-channel single top quark production.}
\label{fig:Feynman}
\end{figure}

Following a suggestion in Ref.~\cite{Yuan:1995sh}, we determine the
partial width $\Gtbw$ of the top quark indirectly from 
the single top $t$-channel ($\ppbar\to tqb+X$) cross section
measurement~\cite{d0_tchan_obs},
assuming that the electroweak coupling in top quark production is
identical to the 
coupling in top quark decay. 
Electroweak single top quark production proceeds via $s$-channel
production and decay of a virtual 
$W$~boson, or through exchange of a virtual $W$~boson in the
$t$-channel~\cite{Willenbrock:1986cr,Yuan:1989tc}. 
As in the decay of top quarks, both processes involve the $Wtb$ vertex
and are therefore proportional to the
partial width $\Gtbw$. 
Since contributions outside the SM have different effects on
the $s$-channel and $t$-channel cross  
sections, the partial width is determined 
focusing on the single most sensitive channel in single top quark
production, the $t$-channel, which is
illustrated in Fig.~\ref{fig:Feynman}. 

From the partial decay width and the
branching fraction $\Btbw$~\cite{rd0}, we form the total decay width:
%
\begin{equation}
\label{eq:tot_width}
\Gamma_t = \frac{\Gtbw}{\Btbw}. 
\end{equation}
%
In addition to the experimental measurements, this method relies on 
the validity of the NLO QCD calculations of the single top quark cross
section and of the top quark partial decay width. In these
calculations only the contributions from SM processes are considered.
Any deviation of the
measured total width from the theoretical prediction would
therefore indicate physics beyond the SM. 
One example is the presence of anomalous form factors in the $Wtb$ vertex,
such as right-handed vector couplings~\cite{Abazov:2008sz}. This would
change the measured 
$t$-channel cross section and therefore the extracted partial width.
It would also modify the measurement of \Btbw\ due to the different event
kinematics which would lead to different event selection efficiencies.
Another example is a charged Higgs boson $H^+$
with a mass $m_{H^+}<m_t-m_b$ which preferably decays via $H^+
\rightarrow c\bar{s}$ as predicted in some supersymmetric extensions
of the SM~\cite{mhdm}.
In this case, the fusion process $H^+ b \rightarrow t$ can
compete with the SM single top quark production ($W^+ b \rightarrow t$),
and the decay $t \rightarrow H^+ b$ can compete with the SM decay
($t \rightarrow W^+ b$). This would enhance the measured $t$-channel
cross section, and affect the measured branching fraction \Btbw.
A third example studied in more detail in this Letter is a new fourth
generation $b'$ quark.

%
To extract the partial width $\Gtbw$, we use the measurement of the inclusive
$t$-channel cross section obtained from data corresponding to
2.3~fb$^{-1}$ of integrated 
luminosity~\cite{d0_tchan_obs}.
Without assuming $\Btbw=1$ as in that publication, the cross section
measurement can be expressed as
%
\begin{equation}
\label{eq:tchan}
\sigma(t\mathrm{-channel})\,\Btbw = 3.14^{+0.94}_{-0.80}~\mathrm{pb}\ .
\end{equation}
%
Given the linear dependence of the cross section on the partial width, we
derive the partial width as
%
\begin{equation}
\label{eq:partwidth}
\Gtbw = \sigma(t\mathrm{-channel}) \ \frac{ \Gtbw_{\rm SM}}
{\sigma(t\mathrm{-channel})_{\rm SM}} \,\,.
\end{equation}
%
For the predicted SM $t$-channel cross section, we use a calculation in NLO 
QCD that yields $\sigma(t\mathrm{-channel})_{\rm  SM} = 2.14 \pm
0.18$~pb~\cite{kidonakis} for $m_t=170$~GeV. 
For the partial width in the SM, we use the NLO
result of ${\Gtbw_{\rm SM}}=1.26$~GeV from Eq.~\ref{eq:smwidth}.
Using Eqs.~\ref{eq:tot_width} and~\ref{eq:partwidth}, the total width
becomes:
%
\begin{equation}
\label{eq:totwidth_derivation}
\Gamma_t = \frac{\sigma(t\mathrm{-channel}) \ \Gtbw_{\rm SM}}
  {\Btbw \ \sigma(t\mathrm{-channel})_{\rm SM}} \,\,.
\end{equation}

The branching fraction $\Btbw$ is determined from our previous studies of 
$\ttbar$ events with different identified $b$ jet multiplicities~\cite{rd0}:
\begin{equation}
\label{eq:R}
\Btbw = 0.962^{+0.068}_{-0.066} ({\rm stat}) \
^{+0.064}_{-0.052} ({\rm syst}) \,\,.
\end{equation}
%
The $\Btbw$ measurement (Eq.~\ref{eq:R}) is used twice: 
to obtain the partial width in Eqs.~\ref{eq:tchan}
and~\ref{eq:partwidth}, and to derive the total width in
Eq.~\ref{eq:totwidth_derivation}. 


The analysis starts with the same Bayesian Neural Network (BNN)
discriminants trained 
to measure the $t$-channel cross section~\cite{d0_tchan_obs} in 24 independent 
analysis channels, separated according to data-taking period, 
lepton flavor ($e$ or $\mu$), jet multiplicity (2, 3 or 4), and number 
of $b$-tagged jets (1 or 2). We then form a Bayesian
probability density~\cite{st_prd} for the partial width based on
Eq.~\ref{eq:partwidth}.  
This is combined with the measurement of $\Btbw$ which is performed
selecting 3 and 4 jets, 
and 0, 1 or 2 $b$-tags for the $e$ and $\mu$ channels.
In combining the probability densities we assume that all the
values of $\Gtbw$ are equiprobable, which corresponds to
assuming a uniform probability density for the $t$-channel cross
section and for $\Gamma_t$.


Systematic uncertainties are treated in the same way as for the combination~\cite{combi} of the CDF~\cite{cdfobs} and D0~\cite{d0obs} single top quark cross section measurements. 
The terms included in the uncertainty calculation are: 
\begin{itemize}
\item Uncertainty on the integrated luminosity of 6.1\%.
\item Uncertainties on modeling the single top quark signal, which
applies only to the $t$-channel cross  
section and includes uncertainties from initial- and final-state radiation, 
scale uncertainties and parton distribution functions ({\sc pdf}s).
\item Uncertainties in the modeling of the $\ttbar$ pair production signal for the \Btbw\ measurement,
which 
include uncertainties from {\sc pdf}s, different event generators and
hadronization 
models. They are correlated with the $\ttbar$ background yield uncertainty
in the $t$-channel measurement.
\item Uncertainties on the background MC simulation, including the $\ttbar$
normalization uncertainty 
in the $t$-channel obtained from theoretical calculations taking into
account the uncertainty 
on $m_t$, and for \Btbw\ the uncertainty on
the $W$+jets and heavy-flavor samples normalization.
\item Detector simulation uncertainty arising from the modeling
of particle identification in MC.
\item Uncertainties arising from the modeling
of the different background sources that are obtained using
data-driven methods. 
\item Uncertainty on $b$-jet identification involving $b$, $c$ and
light-flavor jet tagging rates and the calorimeter response to $b$-jets.
\item Jet energy scale (JES) uncertainty from the calorimeter response
to light jets, 
uncertainties from JES corrections dependent on pseudorapidity and
transverse momentum and
other smaller contributions.
\end{itemize}

All systematic uncertainties of the $t$-channel cross section and the
\Btbw\ measurement are assumed to be either fully correlated or uncorrelated. 
Table~\ref{tab:sys} shows the relative systematic uncertainties
used in the $t$-channel and \Btbw\ measurements, and
displays how the correlations are treated.

\begin{table*}[ht]
\begin{center}
\begin{minipage}{5.5 in}
\begin{ruledtabular}
\begin{tabular}{lccc}
Sources & $\sigma(t$-channel), \% & \Btbw, \%  & Correlations\\  
\hline
Components for normalization      &                &      &     \\
~~Luminosity                            &  6.1         & 0.0  &    \\
~~Single top quark signal modeling      &  3.5--13.6   & 0.0  &    \\
~~Top pair production signal modeling   &  ---         & 1.0  &   X \\
~~Other background from MC              &  15.1        & 0.6  &   X  \\
~~Detector modeling                     &  7.1         & 0.1  &   X  \\
Components for normalization and shape      &      &        &  \\
~~Background from data                  &  13.7--54    &  1.7 & X \\
~~$b$-jet identification                &  2--30       &  6.3 & X \\
~~Jet energy scale                      &  0.1--13.1   &  0.0 &  \\
\end{tabular}
\end{ruledtabular}
\caption[gensys]{
Sources of systematic uncertainties affecting the determination of 
$\Gamma_t$, including sources that affect both the normalization and 
the shape of the final discriminant. 
For some uncertainties we quote the range across the different channels. 
In the $t$-channel cross section measurement the top pair
production modeling uncertainty is included in the ``Other background
from MC'' modeling category. It is taken as fully correlated to the 
``Top pair production signal modeling'' uncertainty in the \Btbw\ measurement. 
The sources are 100\% correlated between the two measurements for
rows with an ``X'' in the correlations column, and uncorrelated otherwise.}
\label{tab:sys}
\end{minipage}
\end{center}
\end{table*}

\begin{figure}[t]
\includegraphics[width=3.2in]{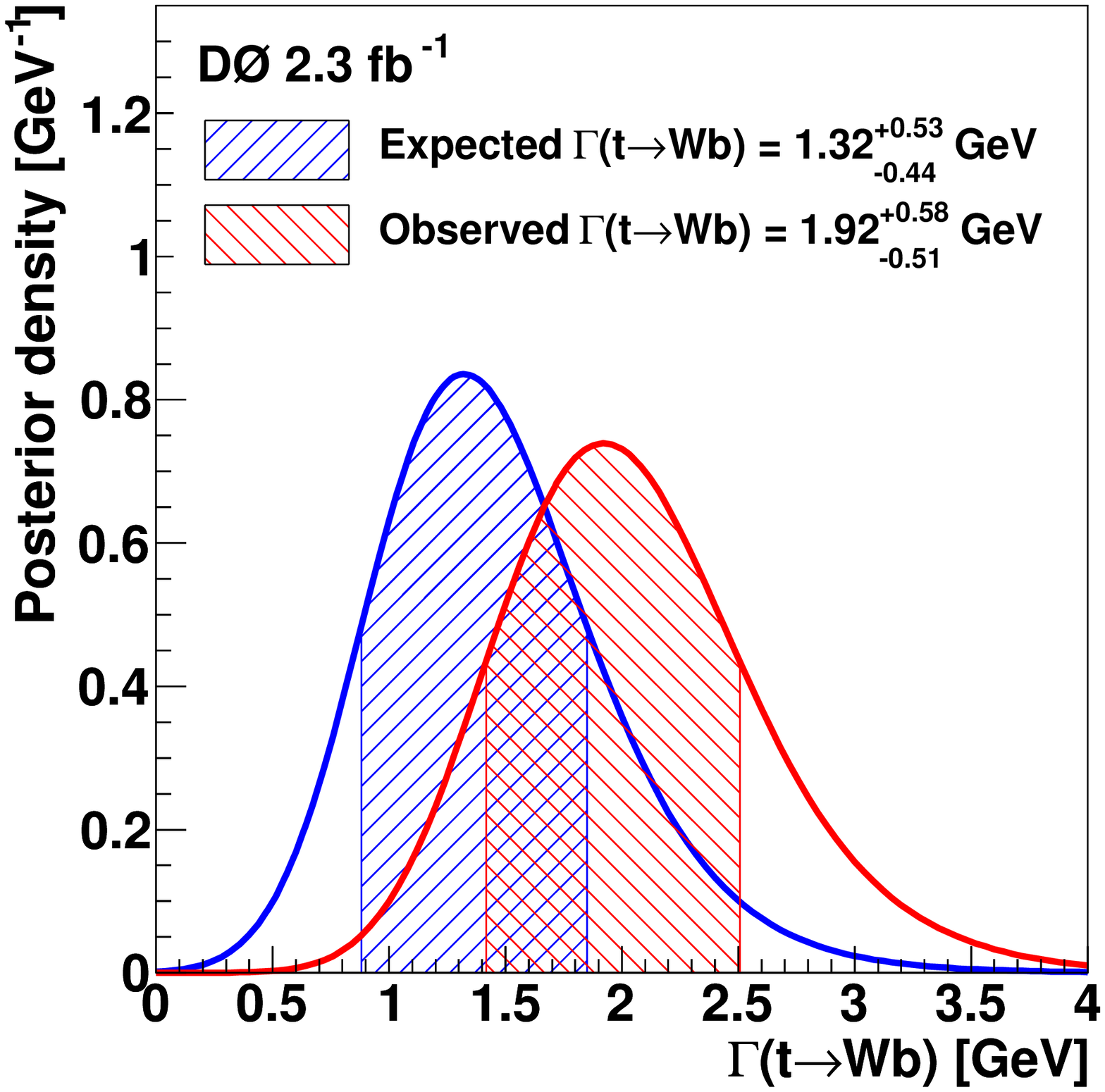}
\vspace{-0.5cm}
\caption{Probability density for the expected and measured partial
width $\Gtbw$. The hatched areas 
represent one standard deviation around the most probable value.} 
\label{fig:tqb_Gbtw_posterior}
\end{figure}

The expected and observed Bayesian probability densities for the partial width
$\Gtbw$ are shown in Fig.~\ref{fig:tqb_Gbtw_posterior}.
%
The most probable value for the partial width is defined by the peak of
the probability density function and corresponds to 
\begin{equation}
\label{eq:partialw_result}
\Gtbw = 1.92^{+0.58}_{-0.51}~\mathrm{GeV}.
\end{equation}

The measurement of the partial width alone can be used to set a lower limit
on the total width. From the observed partial width probability density in 
Fig.~\ref{fig:tqb_Gbtw_posterior}, we obtain that $\Gtbw > 1.21$~GeV at 95\%
C.L. 
This is the lowest value of the partial width 
that bounds 95\% of the area of the probability density.
Since the total width must be larger than the partial width, it also
must satisfy
\begin{equation}
\Gamma_t > 1.21~\mathrm{GeV~at~95\%~C.L.}
\end{equation}
Calculating the lifetime $\tau_t$ as the inverse of the total width,
we determine an upper limit of $\tau_t <
5.4\times10^{-25}$~s. 
Models including an additional chiral-tensorial $Wtb$ coupling leading to
non-SM helicity amplitudes of the  
top quark can be excluded by this result because they predict a
partial width $\Gtbw = 0.66$~GeV~\cite{Nelson}.  

\begin{figure}[!h!tbp]
\includegraphics[width=3.2in]{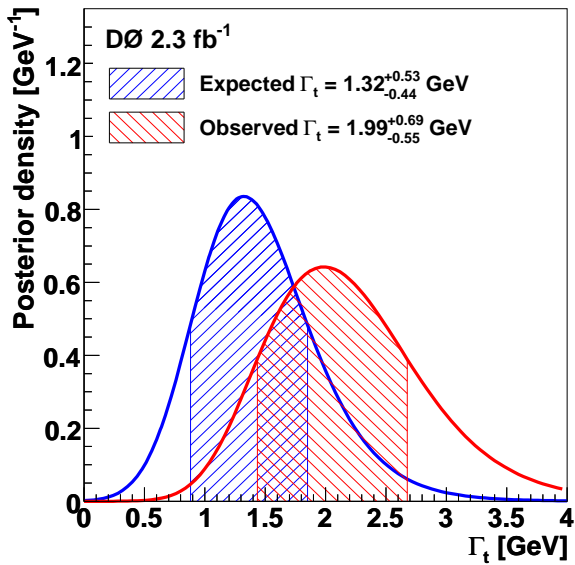}
\vspace{-0.5cm}
\caption{Probability density for the expected and measured
total width $\Gamma_t$.  
The hatched areas represent one standard deviation around the most
probable value.} 
\label{fig:total_Gbtw_posterior}
\end{figure}

Combining the partial width (Eq.~\ref{eq:partialw_result}) 
with $\Btbw$ as in Eq.~\ref{eq:tot_width}, 
we obtain the expected and observed probability densities for the total width 
$\Gamma_t$ shown in Fig.~\ref{fig:total_Gbtw_posterior}. 
%
The total top quark width is found to be
%
\begin{equation}
\Gamma_t = 1.99^{+0.69}_{-0.55}~\mathrm{GeV},
\end{equation}
which can be expressed as a top quark lifetime of 
\begin{equation}
\tau_t = (3.3^{+1.3}_{-0.9})\times10^{-25}~{\rm s}.
\end{equation}

The determination of the top quark width is used to set
constraints on the coupling of a fourth generation $b'$ quark to the
top quark.
Assuming $m_{b'}>m_t-m_W$, a small probability density for the
$b'$ quark in protons and antiprotons, and unitarity of the quark-mixing
matrix, including the fourth quark  
generation ($|V_{tb}|^2+|V_{tb'}|^2=1$, and $|V_{td}|, |V_{ts}| \ll
1$), the
measurement of the total top quark width can be used to extract a limit on
the mixing matrix element $|V_{tb'}|$. Using a flat
prior for $0 \le |V_{tb}| \le 1$ yields $|V_{tb'}|
< 0.63$ at 95\% C.L. This is the first limit
on the $W$~boson coupling to the top quark and a fourth-generation $b'$ quark. 

In summary, we have presented the most precise determination of the width of
the top quark. It is based on the measurement of two quantities, the
partial decay width of the top quark into $Wb$ and the branching
fraction \Btbw. It is assumed that the coupling leading to $t$-channel single top
quark production is identical to the coupling leading to top quark
decay. The total top quark width is determined to be $\Gamma_t =
1.99^{+0.69}_{-0.55}~\mathrm{GeV}$ for $m_t=170$~GeV,
which corresponds to a top quark lifetime of $\tau_t =
(3.3^{+1.3}_{-0.9})\times10^{-25}~{\rm s}$.
In addition, we set the first limit on a fourth-generation $b'$ quark coupling
to the top quark of $|V_{tb'}|
< 0.63$ at 95\% C.L.

\section*{Acknowledgments}
We wish to thank C.-P. Yuan for fruitful discussions regarding this analysis. 
%
We thank the staffs at Fermilab and collaborating institutions,
and acknowledge support from the
DOE and NSF (USA);
CEA and CNRS/IN2P3 (France);
FASI, Rosatom and RFBR (Russia);
CNPq, FAPERJ, FAPESP and FUNDUNESP (Brazil);
DAE and DST (India);
Colciencias (Colombia);
CONACyT (Mexico);
KRF and KOSEF (Korea);
CONICET and UBACyT (Argentina);
FOM (The Netherlands);
STFC and the Royal Society (United Kingdom);
MSMT and GACR (Czech Republic);
CRC Program and NSERC (Canada);
BMBF and DFG (Germany);
SFI (Ireland);
The Swedish Research Council (Sweden);
and
CAS and CNSF (China).

\end{document}